\documentstyle[prl,aps,floats,epsf]{revtex}

\newcommand{\cdfref}[1]{\ignorespaces $^{#1}$}
\newcommand{\DR}{\mbox{$\Delta R$}}
\newcommand{\pt}{\mbox{$p_T$}}
\newcommand{\et}{\mbox{$E_T$}}
\newcommand{\met}{\mbox{$\not\!\!E_T$}}
\newcommand{\metv}{\mbox{$\vec{\not\!\!E}_T$}}

\newcommand{\tanb}{\mbox{$\tan\beta$}}
\newcommand{\gev}{\mbox{${\rm GeV}$}}
\newcommand{\gevc}{\mbox{${\rm GeV}/c$}}
\newcommand{\gevcc}{\mbox{${\rm GeV}/c^2$}}
\newcommand{\tev}{\mbox{${\rm TeV}$}}
\newcommand{\pb}{\mbox{${\rm pb}$}}
\newcommand{\invpb}{\mbox{${\rm pb}^{-1}$}}
\newcommand{\mtop}{\mbox{$M_{top}$}}
\newcommand{\mhpm}{\mbox{$M_{H^\pm}$}}
\newcommand{\hpm}{\mbox{$H^\pm$}}

\begin{document}

\title{%
\vspace*{-25pt}
\hfill Search For Charged Higgs Decays of the Top Quark \hfill 
\raisebox{1.0cm}[0pt]{\makebox[0pt][r]{{\small\sc FERMILAB-PUB-97/058-E}}}
\\ 
Using Hadronic Decays of the Tau Lepton}

\author{%
F.~Abe,\cdfref {17} H.~Akimoto,\cdfref {36} A.~Akopian,\cdfref {31}
M.~G.~Albrow,\cdfref 7 S.~R.~Amendolia,\cdfref {27} D.~Amidei,\cdfref
{20} J.~Antos,\cdfref {33} S.~Aota,\cdfref {36} G.~Apollinari,\cdfref
{31} T.~Asakawa,\cdfref {36}
% ...
\\ (CDF Collaboration) }

\address{%
\begin{center}
\cdfref 1  {Argonne National Laboratory, Argonne, Illinois 60439} \\
\cdfref 2  {Istituto Nazionale di Fisica Nucleare, University of Bologna,
I-40127 Bologna, Italy} \\
\cdfref 3  {Brandeis University, Waltham, Massachusetts 02264} \\
\cdfref 4  {University of California at Los Angeles, Los 
Angeles, California  90024} \\  
\cdfref 5  {University of Chicago, Chicago, Illinois 60638} \\
\cdfref 6  {Duke University, Durham, North Carolina  28708} \\
\cdfref 7  {Fermi National Accelerator Laboratory, Batavia, Illinois 
60510} \\
\cdfref 8  {University of Florida, Gainesville, FL  33611} \\
\cdfref 9  {Laboratori Nazionali di Frascati, Istituto Nazionale di Fisica
               Nucleare, I-00044 Frascati, Italy} \\
\cdfref {10} {University of Geneva, CH-1211 Geneva 4, Switzerland} \\
\cdfref {11} {Harvard University, Cambridge, Massachusetts 02138} \\
\cdfref {12} {Hiroshima University, Higashi-Hiroshima 724, Japan} \\
\cdfref {13} {University of Illinois, Urbana, Illinois 61801} \\
\cdfref {14} {Institute of Particle Physics, McGill University, Montreal 
H3A 2T8, and University of Toronto,\\ Toronto M5S 1A7, Canada} \\
\cdfref {15} {The Johns Hopkins University, Baltimore, Maryland 21218} \\
\cdfref {16} {Universotaet Karlsruhe, 76128 Karlsruhe, Germany} \\
\cdfref {17} {National Laboratory for High Energy Physics (KEK), Tsukuba, 
Ibaraki 315, Japan} \\
\cdfref {18} {Ernest Orlando Lawrence Berkeley National Laboratory, 
Berkeley, California 94720} \\
\cdfref {19} {Massachusetts Institute of Technology, Cambridge,
Massachusetts  02139} \\   
\cdfref {20} {University of Michigan, Ann Arbor, Michigan 48109} \\
\cdfref {21} {Michigan State University, East Lansing, Michigan  48824} \\
\cdfref {22} {University of New Mexico, Albuquerque, New Mexico 87132} \\
\cdfref {23} {The Ohio State University, Columbus, OH 43320} \\
\cdfref {24} {Osaka City University, Osaka 588, Japan} \\
\cdfref {25} {Universita di Padova, Istituto Nazionale di Fisica 
          Nucleare, Sezione di Padova, I-36132 Padova, Italy} \\
\cdfref {26} {University of Pennsylvania, Philadelphia, 
        Pennsylvania 19104} \\   
\cdfref {27} {Istituto Nazionale di Fisica Nucleare, University and Scuola
               Normale Superiore of Pisa, I-56100 Pisa, Italy} \\
\cdfref {28} {University of Pittsburgh, Pittsburgh, Pennsylvania 15270} \\
\cdfref {29} {Purdue University, West Lafayette, Indiana 47907} \\
\cdfref {30} {University of Rochester, Rochester, New York 14628} \\
\cdfref {31} {Rockefeller University, New York, New York 10021} \\
\cdfref {32} {Rutgers University, Piscataway, New Jersey 08854} \\
\cdfref {33} {Academia Sinica, Taipei, Taiwan 11530, Republic of China} \\
\cdfref {34} {Texas A\&M University, College Station, Texas 77843} \\
\cdfref {35} {Texas Tech University, Lubbock, Texas 79409} \\
\cdfref {36} {University of Tsukuba, Tsukuba, Ibaraki 315, Japan} \\
\cdfref {37} {Tufts University, Medford, Massachusetts 02155} \\
\cdfref {38} {University of Wisconsin, Madison, Wisconsin 53806} \\
\cdfref {39} {Yale University, New Haven, Connecticut 06511} \\
\end{center}
}

\date{\today}

\maketitle

\begin{abstract}

This Letter describes a direct search for charged Higgs boson
production in $p\bar{p}$ collisions at $\sqrt{s}=1.8$~\tev\ recorded
by the Collider Detector at Fermilab\@. Two-Higgs-doublet extensions
to the standard model predict the existence of charged Higgs bosons
(\hpm).  In such models, the branching fraction for top quarks ${\cal
B}(t\to H^+ b \to \tau^+\nu\,b)$ can be large.  This search uses the
hadronic decays of the tau lepton in this channel to significantly
extend previous limits on \hpm\ production.

\end{abstract}


\twocolumn

\noindent Many extensions to the standard model (SM), including a
large class of supersymmetric models, have an expanded Higgs sector
containing two Higgs doublets where one doublet couples to the up-type
quarks and neutrinos, and the other couples to the down-type quarks
and charged leptons~\cite{higgstheory}. In these theories, electroweak
symmetry breaking produces five Higgs bosons, three of which are
neutral and two of which are charged.

Recent searches for charged Higgs bosons (\hpm) include analyses
from $p\bar{p}$ collisions at the $Sp\bar{p}S$~\cite{uahiggs} and the
Tevatron~\cite{couyoumtzelis:1996,wang:1994}, from $e^+e^-$ collisions
at CESR~\cite{alam:1995} and at LEP~\cite{LEP}, and from world
averages of the tau lepton branching ratios. An indirect limit from
these averages excludes at 90\% confidence level (C.L.) any charged
Higgs with $\mhpm<1.5\,\tanb$~\gevcc~\cite{LEPew}\ where
\tanb\ is the ratio of the vacuum expectation values of the two Higgs
doublets.

Based on a measurement of the inclusive $b\to s\gamma$ cross section,
CLEO indirectly excludes at 95\% C.L.\ charged Higgs bosons with
$\mhpm\lesssim 244$~\gevcc\ for $\tanb\gtrsim 50$, assuming only a
two-Higgs-doublet extension to the standard
model~\cite{alam:1995}\@. Models with a richer particle structure,
such as supersymmetry, can evade this limit with compensating
destructive interference from particles other than the $W$ and
\hpm~\cite{goto:1996}.

Based on direct searches for charged Higgs pair production, the LEP
experiments exclude at 95\% C.L.\ any charged Higgs
with a mass lower than 44.1~\gevcc~\cite{LEP}.

The Collider Detector at Fermilab (CDF) and D\O\ have
recently established the existence of the top quark via its
semileptonic decays \cite{svxalg,dzerotop}\ using $p\bar{p}$
collisions at $\sqrt{s}=1.8$~\tev\@.  In the analysis presented here,
we search for $t\bar{t}$ events in which the top quarks decay into
charged Higgs bosons. By using a data sample which is five times
larger, improved tau lepton identification, and $b$-quark tagging,
this analysis significantly extends the charged Higgs limits from a
similar, previous CDF search~\cite{couyoumtzelis:1996}.

CDF is a magnetic spectrometer containing tracking detectors,
calorimeters, and muon chambers~\cite{cdfdetector}.  The tracking
detectors lie inside a 1.4~T solenoidal magnetic field.  The central
tracking chamber (CTC) measures the momenta of charged particles over
a pseudorapidity range $|\eta|<1.1$ where $\eta\equiv
-\ln\tan(\theta/2)$~\cite{coordinate}. A silicon vertex detector,
positioned immediately outside the beampipe and inside the CTC,
provides precise charged particle reconstruction and allows
identification of secondary vertices from $b$-quark decays~\cite{svx}.
Electromagnetic and hadronic calorimeters, arranged in a projective
tower geometry, surround the tracking volume and are used to identify
jets, localized clusters of energy, over the range $|\eta|<4.2$. The
presence of neutrinos can be deduced from the missing transverse
energy \met~\cite{metdef}.

This analysis relies on data collected with the \met\ trigger which
nominally requires $\met>35$~\gev\ but is only fully efficient for
$\met\gtrsim 80$~\gev\@.  These data, collected from 1992 to 1995,
represent an integrated luminosity of $100\pm 8$~\invpb.

The ratio \tanb\ determines the dominant decay modes for the \hpm\ 
and top quark. We consider only the region $\tanb\gtrsim 5$ for which
\hpm\ decays to $\tau \nu$ exclusively.  For $\tanb\gtrsim 100$,
both top quarks decay via $t\to Hb$, producing distinctive events with
two tau leptons, two $b$-quarks, and large \met.  For intermediate
\tanb, one or both of the top quarks can decay via $t\to Wb$,
producing events with lower \met\ and fewer tau leptons. To separate
\hpm\ events from background, events in our final sample must have
one of the two following final states.

In the first final state (``$\tau jjX$''), events contain one
hadronically-decaying tau lepton, two jets, and one or more additional
objects.  The other object(s) can be either another lepton (electron,
muon, or tau) or jet.  At least one of the jets must have associated
charged particles that form a displaced vertex indicative of a
$b$-quark decay.

The second final state (``di-tau'') preserves acceptance in the region
where the charged Higgs mass approaches the top quark mass.  In this
case, the $b$-jet energies fall below the jet \et\ requirement,
causing events to fail the $\tau jjX$ requirements.  In the {di-tau}
final state, events contain two energetic, hadronically-decaying tau
leptons that are not opposite in azimuth ($\Delta\phi_{\tau\tau} <
160^\circ$).  To avoid double counting, this category excludes events
passing the $\tau jjX$ requirements.

The electron and muon identification cuts are those used for the top
quark search~\cite{topcuts}.  Identified electrons (muons) must have a
minimum \et\ ($\pt\cdot c$) of 10~\gev.

An iterative algorithm which uses a fixed cone size of
$\DR\equiv\sqrt{(\Delta\phi)^2+(\Delta\eta)^2}=0.4$ finds jets in the
calorimeter~\cite{jetalg}.  Those jets that contain charged particles
that form a displaced secondary vertex~\cite{svxalg} are categorized
as $b$-jets.  Jets identified for this analysis have a minimum
uncorrected \et\ of 10~\gev.

Tau lepton identification begins with a jet.  The
tau lepton candidate must have one or three charged particles in a
$10^\circ$ cone about the jet axis and no additional charged particles
in a cone of $30^\circ$. For counting the number of associated charged
particles, only those with a vertex within 5~cm of the tau
vertex and $\pt>1$~\gevc\ are used.  In addition, the tau lepton
candidate's tracks must have the correct total charge $Q=\pm 1e$ and
its mass $M_\tau$, determined from tracks and electromagnetic
calorimeter energy deposits, must be consistent with that of a tau
lepton ($M_\tau<1.8$~\gevcc).  The \et\ of the calorimeter cluster
associated with the tau lepton candidate must exceed 10~\gev; the
largest $p_T$ of an associated charged particle must exceed 10~\gevc.
The identification algorithm requires $|\eta_\tau|<1$ to maintain good
efficiency for tracking charged particles. Tau lepton identification
efficiency is lower than that for electrons or muons, and the fake
rate from jets is significant.

In the $\tau jjX$ final state, one tau lepton must have $\et>20$~\gev;
any other tau leptons must have $\et>10$~\gev\@.  For the otherwise
less stringent di-tau requirements, we raise the \et\ requirement to
30~\gev\ for both tau leptons.

Both final states require $\met>30$~\gev\@. For events in which a jet
is mismeasured, the \metv\ typically points toward a jet.  To remove
much of this background, we require that events satisfy
$\Delta\phi/deg + \met/\gev>60$ where $\Delta\phi$ is the minimum
angle in azimuth between an identified object in the event and
the \metv.  As events with larger \met\ have lower background, this
cut becomes less severe as \met\ increases.

To reduce contamination from $Z$ boson production, we remove any event
that contains a $\mu^+\mu^-$ or $e^+e^-$ pair with an invariant mass
between 75 and 105~\gevcc.

This set of analysis cuts selects 7 events. All pass the $\tau jjX$
requirements; none passes the di-tau requirements. One event has a tau
lepton, an electron, and two jets; the others have a tau lepton and
three or more jets.  As required, all 7 contain a $b$-tagged jet.

Data samples and Monte Carlo simulations provide the estimate of the
number of expected background events.  The expected number of events
containing jets which imitate tau leptons is determined from
data. Monte Carlo simulations of $W$ and $Z$ plus jet production and
diboson ($WW$, $WZ$, and $ZZ$) production are used to determine the
largest contributions from processes which produce tau leptons.

\begin{table}
\caption{Expected Background and Observed Events\label{ewback}}
\begin{center}

\begin{tabular}{l*{3}{d@{$\pm$}d}}
& \multicolumn{2}{c}{$\tau jjX$} 
& \multicolumn{2}{c}{di-tau} 
& \multicolumn{2}{c}{total} \\
\hline
fake taus           & 5.1  & 1.3  & 0.30 & 0.19 & 5.4  & 1.5  \\
$W +\mbox{}$ jet(s) & \multicolumn{2}{c}{---} & 1.3  & 1.3  & 1.3  & 1.3  \\
$Z +\mbox{}$ jet(s) & \multicolumn{2}{c}{---} & 0.6  & 0.3  & 0.6  & 0.3  \\
$WW$,$WZ$,$ZZ$      & 0.04 & 0.04 & 0.04 & 0.04 & 0.08 & 0.06 \\
\hline
expected     & 5.1  & 1.3  & 2.2  & 1.3  & 7.4  & 2.0  \\
observed & \multicolumn{1}{d}{7} & & \multicolumn{1}{d}{0}
& &\multicolumn{1}{d}{7} & \\
\end{tabular}

\end{center}
\end{table}

Events from QCD and electroweak processes in which a jet
mimics a tau lepton dominate the expected background. Unbiased samples
of jets allow measurement of the rate (parameterized as a function of
jet \et) at which jets imitate a tau lepton.  We replace the normal
tau lepton identification cuts with a function that encodes this fake
rate and apply it event-by-event to the \met\ data sample. This
produces a background estimate, $5.4\pm 1.5$, which is absolutely
normalized and includes any process contributing to the fake
background. In addition to the statistical uncertainty, the quoted
uncertainty includes a 25\% systematic uncertainty on the measured
fake rate. This uncertainty is estimated from differences in the
measured fake rates from the various unbiased jet samples.

$W$ and $Z$ plus jets events are generated with the {\sc vecbos} Monte
Carlo program~\cite{vecbos} with an underlying event added by {\sc
herwig}~\cite{herwig}. This simulation uses the $\langle p_T^2
\rangle$ of the jets for the QCD renormalization and factorization
scales, a minimum \pt\ of 8~\gevc\ for jets, and the {\sc cteq3m}
structure functions~\cite{CTEQ}. Measured cross sections for $W+{\rm
jet(s)}$ and $Z+{\rm jet(s)}$ production provide the normalization of
these Monte Carlo samples~\cite{duker}.  Overall, $1.9\pm 1.3$
background events of this type are expected~\cite{mcfp}.

Diboson production contributes only $0.08\pm 0.06$ events to the
background expectation.  This contribution is determined from an {\sc
isajet}~7.06 Monte Carlo~\cite{isajet}\ which includes tree-level
processes for $WW$, $WZ$, and $ZZ$ production. Table~\ref{ewback}
shows the number of background events from all sources.

To check the background estimation, we compare the number of observed
events without the $b$-tagging requirement (119) to the total number
expected from the various backgrounds ($102\pm 21$).  Moreover, we
compare various kinematic distributions (tau \et, \met, etc.) both
with and without $b$-tagging to the prediction from the sum of the
backgrounds; the agreement is excellent.

\begin{figure}
\epsfxsize=\hsize
\centerline{\epsfbox[20 350 485 695]{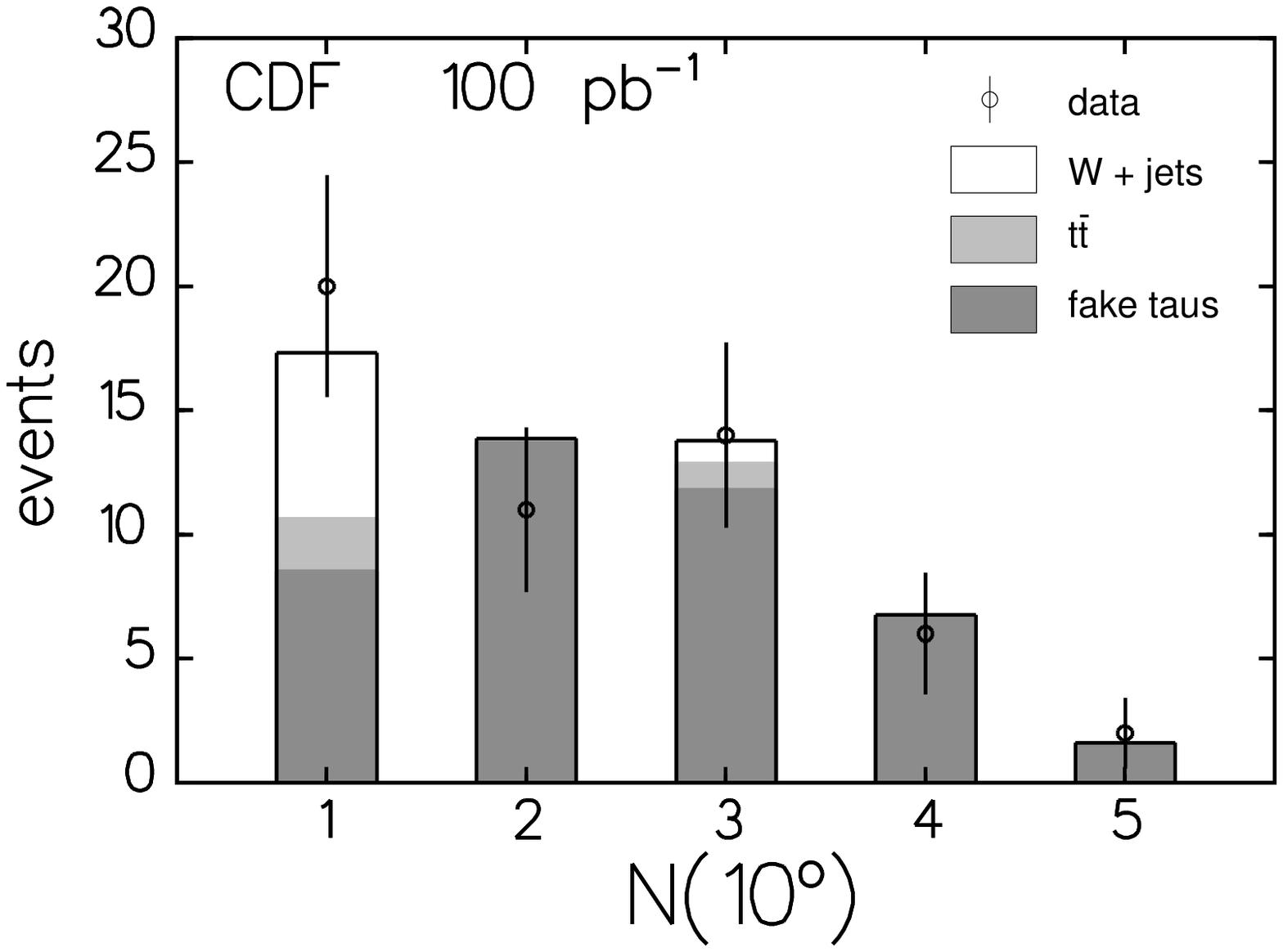}}

\caption{The charged particle multiplicity (in a $10^\circ$ cone) of
tau candidates ($N(10^\circ)$) in the \met\ data sample using cuts
that enhance $W\to\tau\nu+3$ or more jets events.
\label{tauandjets}}
\end{figure}

To verify that the tau lepton identification algorithm works as
expected in events with final states as complex as those from charged
Higgs events, a tau signal from $W\to \tau\nu + 3$ or more jets is
selected from the \met\ data sample. The cuts for doing this differ
slightly from those used for the search. Removing the $b$-tagging
requirement enhances the acceptance, while tightening the \met\
requirement to 40~\gev\ and $\Delta\phi/deg + \met/\gev>75$ reduces
the background in this channel.  Figure~\ref{tauandjets}\ shows the
charged particle multiplicity of tau lepton candidates. (The charge
and charged particle multiplicity requirements for the tau lepton are
not applied.)  The tau signal agrees well with the expectation from
$W$ production once backgrounds are taken into account.

\begin{figure}
\epsfxsize=\hsize
\epsfbox{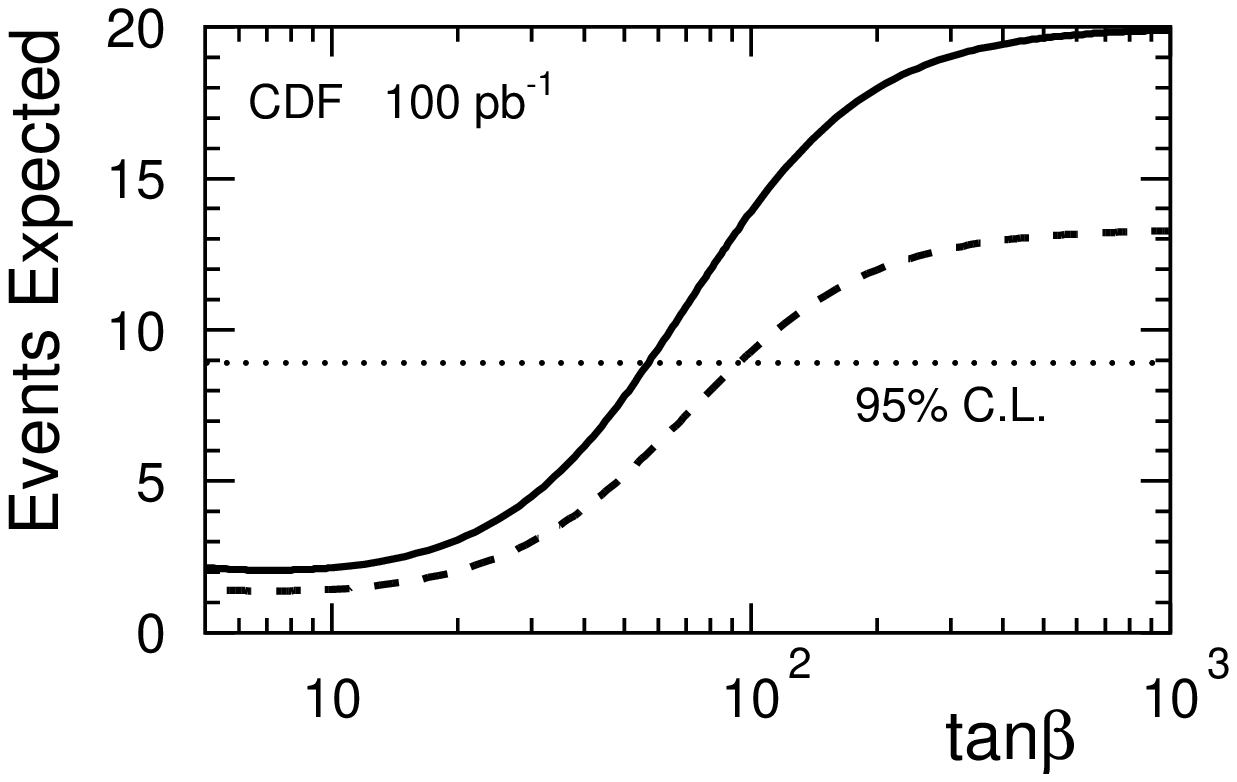}
\caption{Expected number of charged Higgs events for $\mtop=175$
\gevcc, $\mhpm=100$ \gevcc, and $\sigma_{t\bar{t}}=5$ \pb\ (dashed) or
7.5 \pb\ (solid).  Models which predict 8.9 or more expected events
are excluded at 95\% C.L.\ (dotted).\label{hexpect}}
\end{figure}

Based on an {\sc isajet} Monte Carlo simulation, this analysis would
typically retain 2\% of the events in which both top quarks decay into
\hpm\@.  Figure~\ref{hexpect}\ shows the expected number of signal
events as a function of \tanb\ for $\mtop=175$~\gevcc\ and
$\mhpm=100$~\gevcc.  To illustrate the sensitivity to the assumed top
cross section $\sigma_{t\bar{t}}$, the figure shows curves for both
the theoretical value (5~\pb) \cite{topxsec}\ and another value
(7.5~\pb) chosen to be 50\% above the theoretical expectation.  If the
two Higgs doublet model is correct, then any measurement of
$\sigma_{t\bar{t}}$ that assumes the SM decay $t\to Wb$ is an
underestimate of the true $\sigma_{t\bar{t}}$.  The expected
contribution to the {\em signal} when both top quarks decay via the SM
mode is $1.35 \pm 0.12$ events using $\sigma_{t\bar{t}}=5.0$~\pb.

The uncertainty in the fake rate measurement (25\%) dominates the
systematic uncertainty.  Varying parameters of the Monte Carlo
simulations provides estimates of the systematic uncertainties from
the \met\ trigger efficiency (10\%), from inaccurate modeling of gluon
radiation (10\%), and from the overall energy calibration of the
calorimeter (10\%).  The uncertainties from tau identification
efficiency (10\%) and $b$-tagging efficiency (10\%) are determined by
comparing Monte Carlo simulations to various data samples.  The total
systematic error also includes contributions from the integrated
luminosity (8\%) and limited Monte Carlo statistics at the limit
boundary (8\%).  Adding these contributions in quadrature gives a
total systematic uncertainty of 35\%.

Using both final states, 7 events are observed and the expected
background is $7.4\pm 2.0$ events.  This analysis excludes at the 95\%
C.L. any point where the expected number of signal events is 8.9 or
larger. The limit calculation includes the relative systematic
uncertainties on the background and signal~\cite{limit}.

\begin{figure}
\epsfxsize=\hsize
\epsfbox{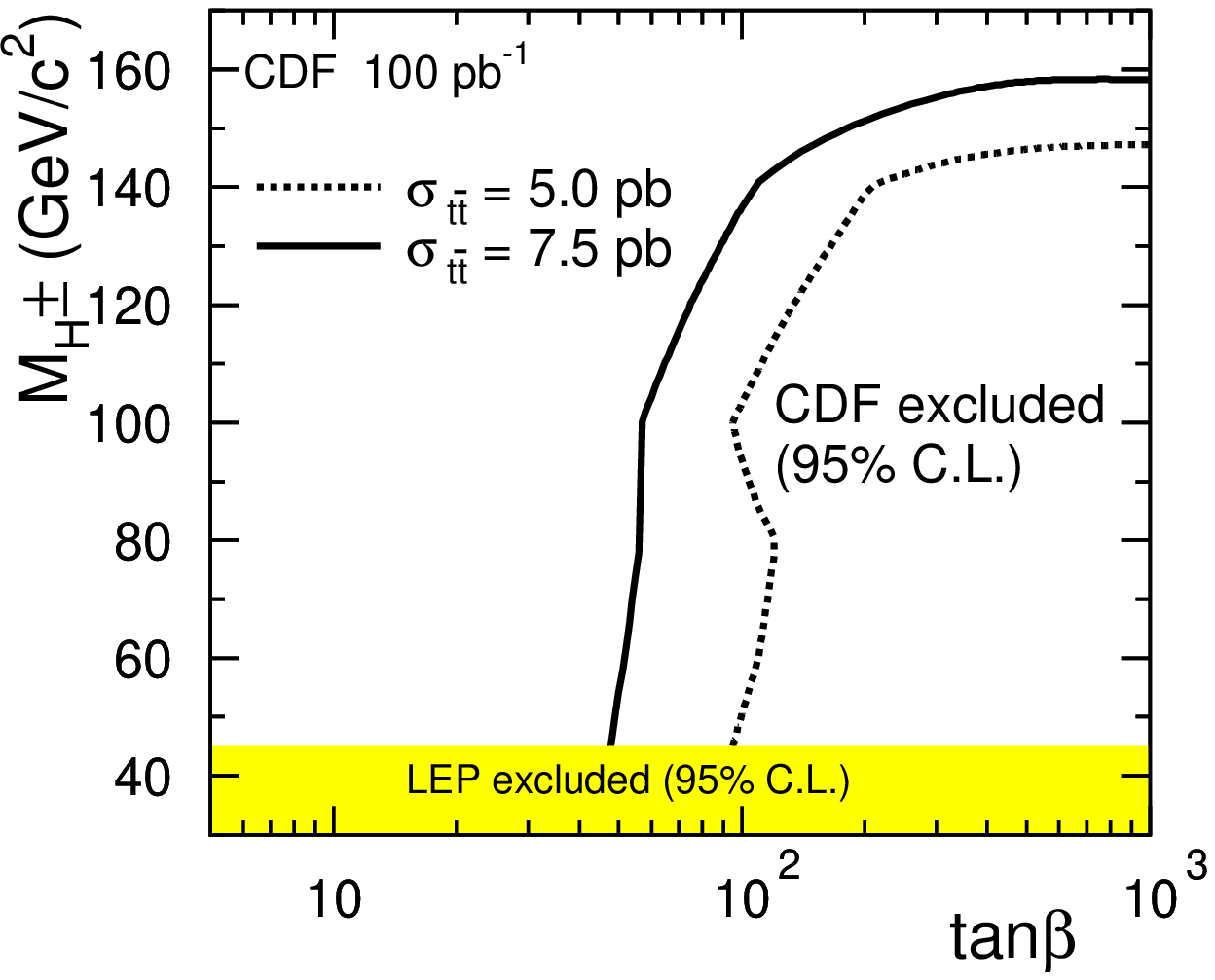}
\caption{Charged Higgs exclusion region for
$\mtop = 175$ \gevcc.\label{limitthree}}
\end{figure}

Figure~\ref{limitthree}\ shows the region excluded by this analysis.
For large \tanb, this analysis excludes charged Higgs bosons with
$\mhpm<147\,\,(158)$~\gevcc\ for a top quark mass of 175~\gevcc\ and
$\sigma_{t\bar{t}}=5.0\,\,(7.5)$~\pb.

\begin{figure}
\epsfxsize=\hsize
\epsfbox{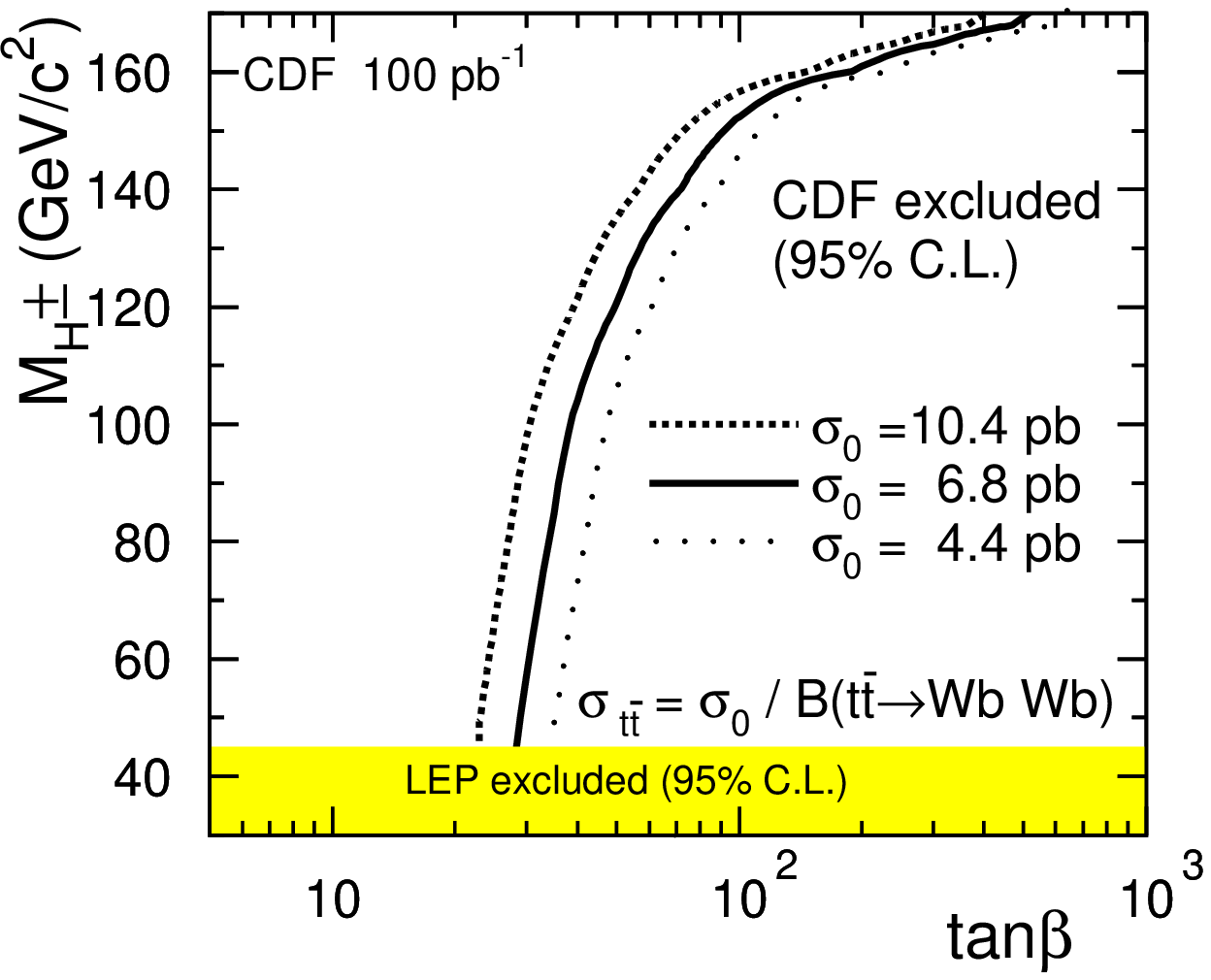}
\caption{Charged Higgs exclusion region for $\mtop = 175$ \gevcc\ 
using $\sigma_{t\bar{t}}=\sigma_0/{\cal B}(t\bar{t}\to
Wb\,Wb)$.\label{limitfour}}
\end{figure}

The top quark discovery provides additional information which can
further restrict \hpm\ production.  To maintain consistency with the
observed top cross section
$\sigma_0=6.8^{+3.6}_{-2.4}$~\pb~\cite{svxalg}, $\sigma_{t\bar{t}}$
must increase at higher \tanb\ to compensate for the lower branching
fraction into the SM mode ${\cal B}(t\bar{t}\to Wb\,Wb)$.
Figure~\ref{limitfour}\ shows the region excluded using this
additional information.

     We thank the Fermilab staff and the technical staffs of the
participating institutions for their vital contributions.  This work was
supported by the U.S. Department of Energy and National Science Foundation;
the Italian Istituto Nazionale di Fisica Nucleare; the Ministry of Education,
Science and Culture of Japan; the Natural Sciences and Engineering Research
Council of Canada; the National Science Council of the Republic of China; 
the A.P. Sloan Foundation; and the Swiss National Science Foundation.


\begin{thebibliography}{10}
\bibitem{coordinate}
The CDF coordinate system defines the polar angle $\theta$ with
respect to the proton beam direction.  The azimuthal angle $\phi$ is
measured in the plane transverse to the beamline.  The transverse
momentum is defined as $\pt\equiv p\sin\theta$; the transverse energy
\et\ is similarly defined.
\bibitem{mcfp}
To avoid double counting background from fake tau leptons, only those
Monte Carlo events in which the reconstructed tau lepton matches a
generated one within $\DR=0.4$ are used.
\end{thebibliography}
\end{document}